\documentclass[twocolumn,amssymb,aps, prb,superscriptaddress,showpacs,reprint]{revtex4-1}
\pdfoutput=1
\usepackage[english]{babel}
\usepackage{braket}
\usepackage{bibentry}
\usepackage{graphicx}
\usepackage{amsmath}
\usepackage{url}
\usepackage{hyperref}
\DeclareMathOperator{\perm}{per}

\begin{document}

\title{An eigenvalue-based method and determinant representations for general integrable XXZ Richardson-Gaudin models}

\author{Pieter~W.\ Claeys}
\email{pieterw.claeys@ugent.be}
\affiliation{Ghent University, Center for Molecular Modeling, Technologiepark 903, 9052 Ghent, Belgium}
\affiliation{Ghent University, Department of Physics and Astronomy, Proeftuinstraat 86, 9000 Ghent, Belgium}
\author{Stijn~De~Baerdemacker}
\affiliation{Ghent University, Center for Molecular Modeling, Technologiepark 903, 9052 Ghent, Belgium}
\affiliation{Ghent University, Department of Physics and Astronomy, Proeftuinstraat 86, 9000 Ghent, Belgium}
\affiliation{Ghent University, Department of Inorganic and Physical Chemistry, Krijgslaan 281 (S3), 9000 Ghent, Belgium}
\author{Mario~Van~Raemdonck}
\affiliation{Ghent University, Center for Molecular Modeling, Technologiepark 903, 9052 Ghent, Belgium}
\affiliation{Ghent University, Department of Physics and Astronomy, Proeftuinstraat 86, 9000 Ghent, Belgium}
\affiliation{Ghent University, Department of Inorganic and Physical Chemistry, Krijgslaan 281 (S3), 9000 Ghent, Belgium}
\author{Dimitri~Van~Neck}
\affiliation{Ghent University, Center for Molecular Modeling, Technologiepark 903, 9052 Ghent, Belgium}
\affiliation{Ghent University, Department of Physics and Astronomy, Proeftuinstraat 86, 9000 Ghent, Belgium}
\date{\today}


\begin{abstract}
We propose an extension of the numerical approach for integrable Richardson-Gaudin models based on a new set of eigenvalue-based variables \cite{faribault_gaudin_2011,el_araby_bethe_2012}. Starting solely from the Gaudin algebra, the approach is generalized towards the full class of XXZ Richardson-Gaudin models. This allows for a fast and robust numerical determination of the spectral properties of these models, avoiding the singularities usually arising at the so-called singular points. We also provide different determinant expressions for the normalization of the Bethe Ansatz states and form factors of local spin operators, opening up possibilities for the study of larger systems, both integrable and non-integrable. These expressions can be written in terms of the new set of variables and generalize the results previously obtained for rational Richardson-Gaudin models \cite{faribault_determinant_2012} and Dicke-Jaynes-Cummings-Gaudin models \cite{tschirhart_algebraic_2014}. Remarkably, these results are independent of the explicit parametrization of the Gaudin algebra, exposing a universality in the properties of Richardson-Gaudin integrable systems deeply linked to the underlying algebraic structure. 
\end{abstract}
\maketitle

\section{Introduction}

Exactly-solvable systems play an important role for the understanding of the physics of quantum many-body systems. They offer an insight into the behaviour of strongly correlated systems in ways that would otherwise be impossible. One class of these systems is the class of Richardson-Gaudin (RG) integrable systems, which can be derived from a generalized Gaudin algebra \cite{dukelsky_colloquium:_2004,ortiz_exactly-solvable_2005}. The pairing model in the reduced Bardeen-Cooper-Schrieffer (BCS) approximation, used to describe superconductivity, has been shown to be RG integrable \cite{cambiaggio_integrability_1997}, as has the $p_x + i p_y$ pairing Hamiltonian \cite{ibanez_exactly_2009,rombouts_quantum_2010}, the central spin model \cite{gaudin_diagonalisation_1976}, factorisable pairing models in heavy nuclei \cite{dukelsky_exactly_2011}, an extended $d+id$ pairing Hamiltonian \cite{marquette_integrability_2013} and several atom-molecule Hamiltonians such as the inhomogeneous Dicke model \cite{dukelsky_exactly_2004}. For these models, diagonalizing the Hamiltonian in an exponentially scaling Hilbert space can be reduced to solving a set of non-linear equations scaling linearly with system size. For most of these systems, computationally inexpensive expressions are also known for several form factors and overlaps, which can be used to investigate observables in systems where the Hilbert space is too large for traditional exact methods. Unfortunately, the Bethe Ansatz or Richardson-Gaudin equations \cite{richardson_exact_1964,gaudin_diagonalisation_1976,dukelsky_class_2001} that need to be solved are highly non-linear and give rise to singularities, making a straightforward numerical solution challenging \cite{richardson_numerical_1966,dominguez_solving_2006}. Several methods have been introduced as a way to resolve this difficulty, such as a change of variables \cite{rombouts_solving_2004,dominguez_solving_2006}, a (pseudo-)deformation of the algebra \cite{de_baerdemacker_richardson-gaudin_2012,van_raemdonck_exact_2014}, or a Heine-Stieltjes connection, reducing the problem to a differential equation \cite{guan_heine-stieltjes_2012}. The ground state energy in the thermodynamic limit has also been obtained by treating the interaction as an effective temperature \cite{pogosov_probabilistic_2012}.

Our method consists of a generalization of the numerical method first proposed by Faribault et al. \cite{faribault_gaudin_2011} for a set of non-degenerate rational RG models, and later extended to degenerate rational models by El Araby, Gritsev and Faribault \cite{el_araby_bethe_2012}. The equations for the Dicke model were independently presented by Babelon and Talalaev \cite{babelon_bethe_2007}. In this method, an alternative set of equations is derived in terms of variables related to the eigenvalues of the constants of motion. One of the advantages of this method is that it offers not only the solutions of the RG equations, but also efficient numerical expressions for form factors and overlaps \cite{faribault_determinant_2012,tschirhart_algebraic_2014}. 
Such expressions for the rational model have already been used to study quantum quenches \cite{faribault_quantum_2009} and decoherence in quantum dots \cite{van_den_berg_competing_2014}, and should facilitate the use of the XXZ model as a variational or projected approach for non-integrable systems, similar to the use of the XXX model in quantum chemistry \cite{johnson_size-consistent_2013,limacher_new_2013,tecmer_assessing_2014}. 

In the following, the results for the rational XXX model are generalized towards the full class of XXZ RG integrable systems and possible applications are discussed.

\section{Richardson-Gaudin models}
\subsection{Definitions}
Unlike its classical counterpart, no single unique definition is known for quantum integrability \cite{caux_remarks_2011}. One class of systems usually denoted integrable is the set of RG integrable systems,  since these support as many (nontrivial) conserved operators commuting with the Hamiltonian as there are degrees of freedom in the system \cite{dukelsky_colloquium:_2004,links_algebraic_2003}. These systems can be diagonalized exactly by means of a Bethe Ansatz wave function, where diagonalizing the full Hamiltonian can be reduced to solving a set of nonlinear coupled equations for the variables in the ansatz \cite{richardson_exact_1964,gaudin_diagonalisation_1976}.

The families of RG integrable systems have their roots in a generalized Gaudin algebra \cite{gaudin_diagonalisation_1976,ortiz_exactly-solvable_2005} based on the $su(2)$ algebra of (quasi-)spin operators \cite{talmi_simple_1993}. Depending on the physical systems, they can either represent spin degrees of freedom, fermion pairs or bosonic/fermionic Schwinger representations. The relevant (quasi-)spin operators can be defined as satisfying the following commutation relations
\begin{eqnarray}
[S_i^0,S^{\dagger}_j]&=&\delta_{ij}S^{\dagger}_i, \qquad  [S_i^0,S_j]=-\delta_{ij}S_i, \nonumber\\ {}[S^{\dagger}_i,S_j]&=&2\delta_{ij}S_i^0,
\end{eqnarray}
where each separate algebra spans a $su(2)_i$ algebra associated with a level $i$ and irreducible representations (irrep) $\ket{d_i,\mu_i}$. For a set of $n$ levels, the RG integrable models are then defined by a set of $n$ mutually commuting operators
\begin{equation}
R_i=S_i^0+g \sum_{k \neq i}^n\left[\frac{1}{2}X_{ik}(S^{\dagger}_i S_k+S_iS^{\dagger}_k)+Z_{ik}S_i^0 S_k^0\right].
\end{equation}
Following Gaudin \cite{gaudin_diagonalisation_1976} and Dukelsky et al. \cite{dukelsky_class_2001}, it is possible to obtain a set of conditions for which the set of operators $R_i$ commute mutually. These are given by
\begin{eqnarray}\label{ga:gaeq}
X_{ij}=-X_{ji}, \qquad Z_{ij}=-Z_{ji}, \\
X_{ij}X_{jk}-X_{ik}(Z_{ij}+Z_{jk})=0,
\end{eqnarray}
the so-called Gaudin equations. They were originally discovered by Gaudin as defining a general class of quadratic Hamiltonians in the spin variables, among which the Gaudin magnet. The derivation by Dukelsky et al. differs from the derivation by Gaudin in the presence of the generator of the Cartan subalgebra $S_i^0$, which however does not influence these conditions. In fact, the Gaudin models are given by Dukelsky's constants of motion in the limit $g \to \infty$ .

Gaudin mentioned three classes of solutions of these equations, where all classes consider $X_{ij}$ and $Z_{ij}$ as odd functions of some real arbitrary parameters $X_{ij}=X(\epsilon_i,\epsilon_j)$. The physical interpretation of these parameters follows from the expression for the Hamiltonian constructed with these parameters.

\begin{enumerate}
\item The rational model
\begin{equation}\label{ga:rat}
X_{ij} = Z_{ij} = \frac{1}{\epsilon_i-\epsilon_j}
\end{equation}
\item The trigonometric model
\begin{equation}\label{ga:trig}
X_{ij} = \frac{1}{\sin(\epsilon_i-\epsilon_j)}, \qquad Z_{ij}=\cot(\epsilon_i-\epsilon_j)
\end{equation}
\item The hyperbolic model
\begin{equation}
X_{ij} = \frac{1}{\sinh(\epsilon_i-\epsilon_j)}, \qquad Z_{ij}=\coth(\epsilon_i-\epsilon_j)
\end{equation}
\end{enumerate}

Here the rational model is also referred to as the XXX-model \cite{dukelsky_class_2001}, indicating that the coefficients in the expression for the constants of motion are identical for all 3 components of the spin. In the same vein, the trigonometric and hyperbolic models are called XXZ-models. Alternative solutions are given by
\begin{eqnarray}\label{ga:richalg}
X_{ij}&=&\frac{\sqrt{1+2\alpha\epsilon_i+\beta \epsilon_i^2}\sqrt{1+2\alpha\epsilon_j+\beta \epsilon_j^2}}{\epsilon_i-\epsilon_j},\nonumber\\
Z_{ij}&=&\frac{1+\alpha(\epsilon_i+\epsilon_j)+\beta \epsilon_i\epsilon_j}{\epsilon_i-\epsilon_j},
\end{eqnarray}
which was proposed by Richardson \cite{richardson_new_2002}, and
\begin{equation}\label{ga:hyp}
X_{ij}=\frac{2\sqrt{\epsilon_i\epsilon_j}}{\epsilon_i-\epsilon_j}, \qquad Z_{ij}=\frac{\epsilon_i+\epsilon_j}{\epsilon_i-\epsilon_j},
\end{equation}
which is a reparametrization of the hyperbolic model \cite{ortiz_exactly-solvable_2005}. 
\subsection{Diagonalizing integrable Hamiltonians}

Any general RG integrable Hamiltonian can be written as a linear combination of the constants of motion
\begin{equation}\label{RG:hamiltonian}
\hat{H}=\sum_{i=1}^n \eta_i R_i,
\end{equation}
resulting in a wide variety of systems. The reduced BCS Hamiltonian can be found from the rational model \cite{cambiaggio_integrability_1997}, the $p_x+ip_y$ pairing Hamiltonian can be derived from the hyperbolic model \cite{ibanez_exactly_2009,rombouts_quantum_2010}, and the central spin model Hamiltonian is identical to one of the constants of motion of the rational model \cite{gaudin_diagonalisation_1976}. By introducing a bosonic degree of freedom, the inhomogeneous Dicke model can be found as a limiting case of the XXZ model \cite{dukelsky_exactly_2004}.

Since all constants of motion commute mutually, they also commute with any Hamiltonian that can be written as Eq. (\ref{RG:hamiltonian}), so this reduces the diagonalization of the Hamiltonian to the diagonalization of one of the constants of motion. It can be shown that the eigenstates are given by a Bethe Ansatz
\begin{equation}\label{rg:bas}
\ket{\psi_N}=\prod_{\alpha=1}^N\left(\sum_{i=1}^n X_{i\alpha}S^{\dagger}_i\right)\ket{\theta},
\end{equation}
with $N$ the number of excitations \footnote{With excitations we denote spin-up flips for magnetic (spin) systems, and the number of pairs for pairing (quasispin) systems.} and the vacuum state $\ket{\theta}=\otimes_{i=1}^n \ket{d_i,-d_i}$ the tensor product of the lowest-weight irreps of each $su(2)_i$ copy, provided the RG equations
\begin{equation}\label{ga:RGeq}
1+g\sum_{i=1}^n Z_{i\alpha}d_i-g\sum_{\beta \neq \alpha}^N Z_{\beta\alpha}=0, \ \  \forall \alpha=1\dots N,
\end{equation}
are satisfied, where the algebra has been extended to $X_{i\alpha}=X(\epsilon_i,x_{\alpha})$ and $Z_{i\alpha}=Z(\epsilon_i,x_{\alpha})$ by introducing a set of $N$ parameters $\{x_{\alpha}\}$. These parameters, also known as the RG variables or rapidities, fix the wave function and need to be determined from the RG equations (\ref{ga:RGeq}). There are multiple strategies to derive this result. Richardson obtained these equations for the reduced level-independent BCS model starting from a variational approach \cite{richardson_restricted_1963,richardson_exact_1964}, Gaudin started from integrability constraints \cite{gaudin_diagonalisation_1976}, Zhou et al. derived these results starting from the Algebraic Bethe Ansatz \cite{zhou_superconducting_2002} and Ortiz et al. used a commutator scheme based on the Gaudin algebra \cite{ortiz_exactly-solvable_2005}.

Once the RG equations have been solved and the rapidities determined, the eigenvalues of the constants of motion $R_i$ can be evaluated as
\begin{equation}
r_i=d_i\left(-1+g\sum_{k\neq i}^nZ_{ik}d_k+g\sum_{\beta=1}^N Z_{\beta i}\right), \ \ \forall i=1 \dots n.
\end{equation}
Although the set of RG equations seems to be linear in the elements $Z_{i\alpha}$ and $Z_{\alpha\beta}$, these variables are still coupled through the Gaudin equations, leading to a set of coupled and nonlinear equations. As an example, the RG equations for the doubly-degenerate ($d_i=1/2, \forall i$) XXX model are given by
\begin{equation}\label{rg:bcs}
1+\frac{g}{2}\sum_{i=1}^n \frac{1}{\epsilon_i-x_{\alpha}} - g \sum_{\beta \neq \alpha}^N \frac{1}{x_{\beta}-x_{\alpha}}=0, \ \ \forall \alpha=1 \dots N.
\end{equation}
These equations need to be solved for the set of rapidities $\{x_{\alpha}\}$ and are coupled and highly nonlinear. It has been shown that singular points arise in these equations at certain values of the coupling constant, where multiple rapidities $x_{\alpha}$ equal one of the single-particle levels $\epsilon_i$ \cite{richardson_numerical_1966,rombouts_solving_2004,dominguez_solving_2006}. It can be readily seen that both the second term and the third term in the RG equations contain diverging poles, but it can also be shown that these singularities cancel exactly. Unfortunately, the behaviour arising at these so-called singular points hampers straightforward solutions of these equations, so involved numerical methods have to be found to circumvent these points.

\section{An eigenvalue-based numerical method}
\subsection{Doubly degenerate models}
For the XXX spin-$1/2$ model ($d_i=1/2, \forall i$), it is possible to introduce a new set of variables 
\begin{equation}
\Lambda_i \equiv \Lambda(\epsilon_i)=\sum_{\alpha=1}^N\frac{1}{\epsilon_i-x_{\alpha}},
\end{equation}
circumventing the singular points in the Richardson-Gaudin equations \cite{faribault_gaudin_2011}. A set of equations equivalent to the RG equations can be found for these variables, however void of singular behaviour. As an illustration, the set of RG equations for the doubly-degenerate XXX model (Eq. \ref{rg:bcs}) are isomorphic to
the set of quadratic equations\cite{faribault_gaudin_2011}
\begin{equation}
\Lambda_i^2=-\frac{2}{g}\Lambda_i+\sum_{j \neq i}^n \frac{\Lambda_j-\Lambda_i}{\epsilon_j-\epsilon_i}, \ \  \forall i=1 \dots n.
\end{equation}
A straightforward numerical solution of these equations can easily be implemented and no singular behaviour will arise since no variables occur in the denominator.
We wish to extend these results to the class of more general XXZ models, relying only on the Gaudin algebra and not on an explicit rational expression of $X$ and $Z$. Define
\begin{equation}
\Lambda_i \equiv \sum_{\alpha=1}^NZ_{i\alpha}=\sum_{\alpha=1}^NZ(\epsilon_i,x_{\alpha}),
\end{equation}
which reduces to the previously-introduced variables in the XXX model. It is interesting to note that the eigenvalue $r_i$ of the constant of motion $R_i$ is related to the variable $\Lambda_i$ as
\begin{equation}
r_i=d_i\left(-1-g\Lambda_i+g\sum_{k\neq i}^n Z_{ik}d_k\right),
\end{equation}
which has led to the denomination eigenvalue-based variables.

In the following, we will start from the Richardson-Gaudin equations for spin-1/2 systems
\begin{equation}
1+\frac{g}{2}\sum_{i=1}^nZ_{i\alpha} = g \sum_{\beta \neq \alpha}^NZ_{\beta\alpha}, \qquad \forall \alpha=1 \dots N,
\end{equation}
and several properties that can be derived from the Gaudin algebra, as previously obtained by Ortiz et al. \cite{ortiz_exactly-solvable_2005}. Firstly, it has been shown that
\begin{equation}
(X_{ij})^2-(Z_{ij})^2=\Gamma, \qquad \forall i \neq j,
\end{equation}
where $\Gamma$ is a constant. The rational model corresponds to $\Gamma=0$, while positive and negative $\Gamma$ result in the trigonometric and hyperbolic models respectively. It also follows from the Gaudin algebra that
\begin{equation}\label{ga:Zeq}
Z_{ij}Z_{jk} - Z_{ik}(Z_{ij}+Z_{jk})=\Gamma, \qquad \forall i \neq j \neq k \neq i,
\end{equation}
which is a generalization of the Gaudin equations for the rational model, where $X_{ij}=Z_{ij}$ and $\Gamma=0$. Using only these equations, it can straightforwardly be shown that
\begin{equation}\label{ebv:eq}
\Lambda_i^2=N(n-N)\Gamma-\frac{2}{g}\Lambda_i+\sum_{j \neq i}^n Z_{ji}(\Lambda_j-\Lambda_i), \ \ \  \forall i=1 \dots n
\end{equation}
with $N$ the total number of excitations and $n$ the number of single-particle levels. The full derivation can be found in Appendix \ref{appendixA}. Unlike the case of the rational model, these equations depend explicitly on the number of excitations $N$. When solving these equations numerically, it was found that the total number of solutions exceeds the dimension of the Hilbert space for $N$ excitations distributed over $n$ levels. Therefore, these equations necessarily support unphysical solutions, not corresponding to any eigenstate, implying that this new set of equations is not equivalent to the original set of RG equations (Eq. \ref{ga:RGeq}). In order to obtain a set of equations equivalent to the original equations, additional constraints for the $\Lambda_i$ are needed.

It is clear from the $\Gamma=0$ case (XXX) that the new set of equations can not distinguish between the different excitation sectors $N$. This can be imposed by noting that the sum of all constants of motion is given by the operator counting the number of excitations
\begin{equation}
\sum_{i=1}^nR_i=\sum_{i=1}^n S_i^0.
\end{equation}
Writing out the eigenvalues of the constants of motion in the new variables results in
\begin{equation}\label{ebv:sum}
-\frac{g}{2}\sum_{i=1}^n \Lambda_i=N.
\end{equation}
In the following section it will be shown that the full set of equations (\ref{ebv:eq}) and (\ref{ebv:sum}) has as many solutions as the dimension of the Hilbert space, so introducing this additional equation leads to a system of equations fully equivalent to the original set of RG equations. These eigenvalue-based equations do not show singular behaviour and can easily be solved numerically, as will be shown in the following.

\subsection{The weak-coupling limit}
In order to obtain some insight in the behaviour of the solutions of these equations, an approximate solution can be found in the weak-coupling limit (small $|g|$). This can be done by proposing a series expansion in $g$ for the solutions of these equations and solving the equations at each order.
For small $g$, a series expansion of $\Lambda_i$ in $g$ can be proposed up to $\mathcal{O}(g)$, keeping only the two dominant terms
\begin{equation}
\Lambda_i = \frac{\lambda_i^{(-1)}}{g}+\lambda_i^{(0)}+\mathcal{O}(g).
\end{equation}
By plugging this expansion in the equations, we obtain
\begin{eqnarray}
&\frac{1}{g^2}\lambda_i^{(-1)}\left(\lambda_i^{(-1)}+2\right)\nonumber\\
&+\frac{1}{g}\left(2\lambda_i^{(0)}(1+\lambda_i^{(-1)})-\sum_{j \neq i}^nZ_{ji}\left[\lambda_j^{(-1)}-\lambda_i^{(-1)}\right]\right)
\nonumber\\
&+\mathcal{O}(g^0)=0.
\end{eqnarray}
This results in
\begin{equation}
\lambda_i^{(-1)}=0 \ \  \textrm{or} \  -2
\end{equation}
and
\begin{equation}
\lambda_i^{(0)}=\frac{1}{2\lambda_i^{(-1)}+2}\sum_{j \neq i}^n Z_{ji}\left(\lambda_j^{(-1)}-\lambda_i^{(-1)}\right).
\end{equation}
In order to satisfy Eq. (\ref{ebv:sum}), the number of dominant terms different from $0$ has to equal the number of excitations $N$, resulting in $\binom{n}{N}$ solutions. The total number of solutions then equals the dimension of the Hilbert space for $N$ excitations distributed over $n$ doubly degenerate levels. Any solution in the weak-coupling limit can be adiabatically connected to a solution for arbitrary coupling, indicating that all possible solutions are always found and no unphysical solutions are present.

This series expansion can also be connected to the series expansion obtained for the rapidities \cite{ortiz_exactly-solvable_2005}. In the limit $g \to 0$ we obtain a noninteracting model, where the rapidities $x_{\alpha}$ converge to the parameters $\epsilon_i$, depending on the corresponding distribution of excitations over energy levels. A rapidity converging to $\epsilon_i$ then corresponds to an excited level $i$ in the noninteracting limit. For finite but small $g$, dominant corrections of $\mathcal{O}(g)$ are present, which are proportional to the roots of orthogonal polynomials via a Heine-Stieltjes connection \cite{stieltjes_theoreme_1885,sriram_shastry_solution_2001}. For $x_{\alpha}$ converging to $\epsilon_i$, this results in $Z_{i\alpha}=Z(\epsilon_i,x_{\alpha})$ diverging as $1/g$ in the weak-coupling limit, where the proportionality factor can be found to be $-2$ from the Heine-Stieltjes connection for $d_i=1/2$. For all other levels $j \neq i$, $Z_{j\alpha}$ remains finite and the dominant order is $\mathcal{O}(g^0)$.

A dominant order of $\mathcal{O}(1/g)$ in $\Lambda_i$ then corresponds to an excited state $i$ in the noninteracting limit, while a dominant order of $\mathcal{O}(g^0)$ results in an non-excited state $i$. This behaviour can be generalized through a connection of $\Lambda_i$ to occupation numbers, as will be shown in the section about form factors. Note that the divergence in $g \to 0$ can pose numerical problems, so this suggests the use of $g\Lambda_i$ as variables instead of $\Lambda_i$.

\subsection{Solving the equations}
Due to the similarity of Eq. (\ref{ebv:eq}) to the set of equations found for the rational model \cite{faribault_gaudin_2011}, it is straightforward to extend the solution method for the rational model to our equations. General sets of nonlinear equations have to be solved by an iterative approach starting from an initial guess, such as the Newton-Raphson method. This method converges quadratically to the solution if the initial guess lies in the basin of attraction. An efficient numerical approach can be implemented based on this method once we have access to a sufficiently good initial guess for the solution. 

As in [\onlinecite{faribault_gaudin_2011}], we start from an approximation of the solution in the weak-coupling limit ($|g|\ll$). A solution at any value of the coupling constant can then be found by adiabatically varying $g$ starting from the weak-coupling limit and using the solution at the previous step as the starting point for an iterative solution at the current step. This initial guess can be improved by using a Taylor expansion of the solutions at the previous step,
\begin{equation}
\Lambda_i(g+\delta g) \approx \Lambda_i(g)+\delta g \frac{\partial \Lambda_i}{\partial g}
\end{equation}
since the derivatives of the $\Lambda_i$ can be found by solving a linear system once the set of $\Lambda_i$ is known, similar to the procedure followed in [\onlinecite{faribault_gaudin_2011}]. The equations for the derivatives can be found by deriving the set of equations to $g$, leading to 
\begin{equation}
\Lambda_i \frac{\partial \Lambda_i}{\partial g}=\frac{\Lambda_i}{g^2}-\frac{1}{g}\frac{\partial \Lambda_i}{\partial g}+\frac{1}{2}\sum_{j \neq i}^n Z_{ji}\left(\frac{\partial \Lambda_j}{\partial g}-\frac{\partial \Lambda_i}{\partial g}\right).
\end{equation}
By taking the higher derivatives of the original set of equations, linear equations can be found for higher order derivatives of $\Lambda_i$, which allows a Taylor expansion up to arbitrary order. For an efficient numerical implementation, combining the Newton-Raphson method with a Taylor approximation up to first order already offers a remarkable increase in speed.

\subsection{Inverting the transformation}
Although many properties of the XXZ systems can be written in terms of the eigenvalue-based variables $\Lambda_i$, as will be shown in the following section, it is still necessary to obtain the rapidities from these variables for the evaluation of e.g. certain form factors. For the rational model, Slavnov's determinant formula \cite{slavnov_calculation_1989} is the building block of many such expressions, and this formula relies on the construction of a matrix expressed in the rapidities. If we wish to obtain a general formalism, it should be possible to obtain the rapidities from the eigenvalue-based variables for any XXZ model.

In order to be as general as possible, we propose a method to obtain the rapidities that does not rely on the explicit expression for $X$ and $Z$ except in the very last step. We introduce an auxiliary index $r$, corresponding to an auxiliary single-particle level or an additional uncoupled rapidity $\epsilon_r$, such that the Gaudin algebra can be extended. Following Ortiz et al. \cite{ortiz_exactly-solvable_2005} and Eq. (\ref{ga:Zeq}), each $Z_{i\alpha}$ can be written as
\begin{equation}\label{var:parZ}
Z_{i\alpha}=\frac{\Gamma+Z_{ri}Z_{r \alpha}}{Z_{ri}-Z_{r\alpha}}.
\end{equation}
Once an explicit expression for $Z_{ij}$ is known, $Z_{ri}=Z(\epsilon_r,\epsilon_i)$ can easily be calculated after choosing $\epsilon_r$. Instead of solving for the rapidities $\{x_{\alpha}\}$, it is now possible to find a transformation that allows us to determine the set of Gaudin algebra elements $\{Z_{r\alpha}\}$. If these are known, the rapidities can always be found by solving each separate equation $Z_{r\alpha}=Z(\epsilon_r,x_{\alpha})$ for $x_{\alpha}$. 

This expression can be used to show that
\begin{eqnarray}\label{inv:lam}
\Lambda_i&=&\sum_{\alpha=1}^NZ_{i\alpha}=\sum_{\alpha=1}^N\frac{\Gamma+Z_{ri}Z_{r \alpha}}{Z_{ri}-Z_{r\alpha}}\nonumber\\
&=&-NZ_{ri}+(\Gamma+Z_{ri}^2)\sum_{\alpha=1}^N\frac{1}{Z_{ri}-Z_{r\alpha}}.
\end{eqnarray}
We can now define a polynomial with the full set ($\alpha=1 \dots N$) of $Z_{r\alpha}$ as roots
\begin{equation}
P(z)=\prod_{\alpha=1}^N(z-Z_{r\alpha})=\sum_{m=0}^N P_{N-m}z^m,
\end{equation}
with $P_0=1$. Once the coefficients $P_{N-m}$ are known, the roots of this polynomial can be determined to find the variables. This method arises naturally for different problems in the theory of integrable systems, such as the Heine-Stieltjes connection \cite{stieltjes_theoreme_1885,sriram_shastry_solution_2001}, the numerical methods by Guan et al. \cite{guan_heine-stieltjes_2012} and Rombouts et al. \cite{rombouts_solving_2004}, and the weak-coupling limit in RG models \cite{ortiz_exactly-solvable_2005}.

The definition of $P(z)$ can now be used to consider
\begin{equation}
\frac{P'(z)}{P(z)}=\frac{\sum_{m=0}^N m P_{N-m}z^{m-1}}{\sum_{m=0}^N P_{N-m}z^m}=\sum_{\alpha=1}^N\frac{1}{z-Z_{r\alpha}}.
\end{equation}
Evaluating this equality in $z=Z_{ri}$, we find that
\begin{equation}
\Lambda_i=-N Z_{ri}+(\Gamma+Z_{ri}^2)\frac{P'(Z_{ri})}{P(Z_{ri})}
\end{equation}
resulting in
\begin{equation}
\frac{P'(Z_{ri})}{P(Z_{ri})}=\sum_{\alpha=1}^N\frac{1}{Z_{ri}-Z_{r\alpha}}=\frac{\Lambda_i+N Z_{ri}}{\Gamma+Z_{ri}^2}.
\end{equation}
The coefficients of the polynomial $P(z)$ can then be found by solving a linear problem
\begin{eqnarray}
\sum_{m=0}^{N-1}P_{N-m}&&\left[m\Gamma Z_{ri}^{m-1}-\Lambda_iZ_{ri}^m+(m-N)Z_{ri}^{m+1}\right]\nonumber\\
&&=\Lambda_i Z_{ri}^N-N\Gamma Z_{ri}^{N-1}.
\end{eqnarray}
We obtain $n$ equations for $N<n$ variables, but if the set of $\Lambda_i$ variables can be written as Eq. (\ref{inv:lam}), these are linearly dependent and we are free to choose $N$ equations from the full set and solve these.
Once these coefficients are known, the variables $Z_{r\alpha}$ can be determined using a root-finding algorithm such as Laguerre's method. Although easy to implement, this method has the disadvantage of being prone to numerical errors. Indeed, it is well-known that the roots of a polynomial are highly sensitive to changes in the coefficients, sometimes even changing the solutions at the qualitative level (a set of complex conjugate roots can be found numerically instead of two separate real roots). These considerations are not pressing for a limited number of excitations, but become more and more important for an increasing number of excitations. 

This problem was also encountered for the rational model \cite{faribault_gaudin_2011}, after which an improved method was proposed by decomposing $P(z)$ in the basis of Lagrange polynomials \cite{el_araby_bethe_2012}. This led to an increased accuracy and allowed the method to tackle problems with a large number of excitations (up to a few hundreds). It is expected that these results can be generalized towards our problem because of the similarity of all necessary equations. We refer the reader to [\onlinecite{el_araby_bethe_2012}] for a detailed analysis of the method.

\begin{figure*}[htb!]                      
 \begin{center}
 \includegraphics[width=\textwidth]{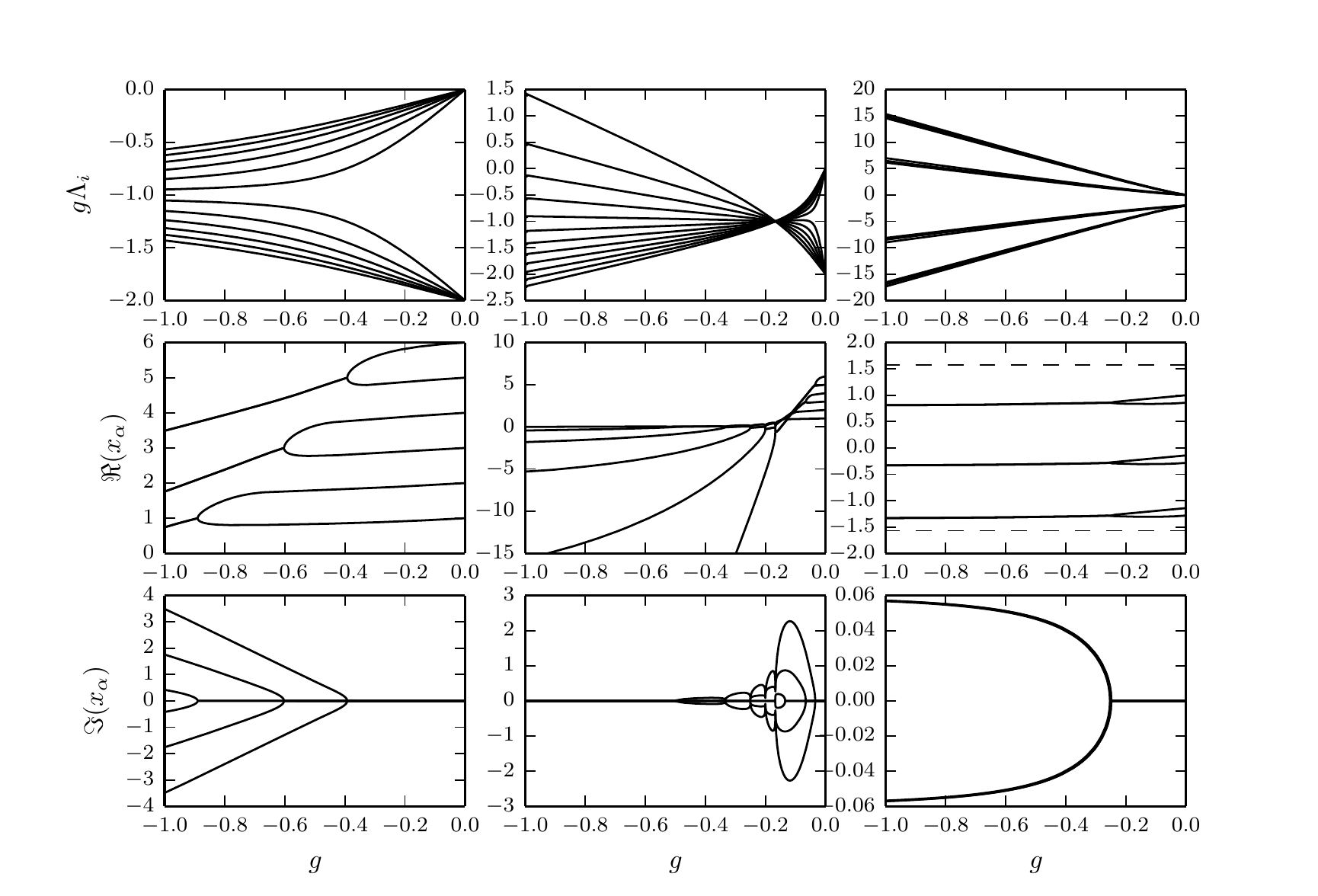}
 \caption{Evolution of the variables for the ground state of a 'picket-fence' model with 6 excitations in 12 doubly-degenerate levels $[1.,2.,3., \dots ,12.]$. The rational, the hyperbolic and the trigonometric models are organized in columns, while the rows depict the set of eigenvalue-based variables $g\Lambda_i$ and the real and imaginary part of the rapidities as a function of the coupling constant $g=0 \dots -1$. The used parametrizations are given by Eqs. (\ref{ga:rat}), (\ref{ga:hyp}) and (\ref{ga:trig}) respectively. Note the Moore-Read point at $g=-2/n=-0.167$ for the hyperbolic model \cite{rombouts_quantum_2010}, where all rapidities condense to zero, and the eigenvalue-based variables all become equal to $-1$. Due to the periodicity of the trigonometric model, the real parts of the rapidities lie within the interval $[-\pi/2,+\pi/2]$, as marked in the figure. \label{fig:models}}
 \end{center}
\end{figure*}

As an illustration, the eigenvalue-based variables and the related rapidities have been plotted for the different models in Figure \ref{fig:models} as a function of the coupling constant. For each model, the levels $\epsilon_i$ are given by a picket-fence model \cite{hirsch_fully_2002} with equal level spacing: $\epsilon_i=i, \forall i$. For each system singular points occur, where multiple real rapidities combine and continue as a pair of complex conjugate variables. As is clear from Figure \ref{fig:models}, the RG equations (Eq. \ref{ga:RGeq}) become highly singular, whereas the eigenvalue-based variables vary very smoothly.

\subsection{Degenerate models}
The discussion has been limited to $d_i=1/2$ models only thus far. This is sufficient for the majority of interacting (quasi-)spin systems, however in situations with higher symmetries arbitrary degeneracies $d_i > 1/2$ may occur. For the rational model, the set of equations for the eigenvalue-based variables were derived starting from a differential equation, where each equation for $\Lambda_i$ is obtained by evaluating this equation at a different level $\epsilon_i$. For degenerate levels, additional equations could be obtained by taking derivatives of the differential equation and evaluating these at the different values $\epsilon_i$ \cite{faribault_gaudin_2011,el_araby_bethe_2012}. This necessitated the introduction of new variables
\begin{equation}
\Lambda_i^{(n)} \sim \sum_{\alpha=1}^N \frac{1}{(\epsilon_i-x_{\alpha})^n},
\end{equation}
which are highly reminiscent of the set of variables introduced by Rombouts et al. \cite{rombouts_solving_2004}. The total number of variables is determined by the number of single-particle levels weighted by their degeneracies. For each single particle level $i$ with degeneracy $2d_i+1$, taking the first $2d_i-1$ derivatives of the differential equation and evaluating each equation at $\epsilon_i$ then results in a closed set of equations for the set of variables $\Lambda_i,\Lambda_{i}^{(2)},\dots,\Lambda_{i}^{(2d_i)}, \forall i$.

The outlined procedure is slightly more involved for the class of XXZ RG models, however it is analogous to the rational case, and remains therefore tractable. The main idea is identical: starting from a continuous representation of the equations it is possible to obtain any number of equations for the total set of variables. The continuous representation of the variables is given by
\begin{equation}
\Lambda(z)=\sum_{\alpha=1}^NZ(z,x_{\alpha}),
\end{equation}
where $\Lambda(\epsilon_i)\equiv\Lambda_i$. A similar derivation as for the $d_i=1/2$ model in Appendix \ref{appendixA} results in a continuous equation
\begin{eqnarray}\label{deg:conteq}
[\Lambda(z)]^2&=&\Gamma N\left(1-N+2\sum_{j \neq i}d_j\right)-\frac{2}{g}\Lambda(z)\nonumber\\
&+&2\sum_{j \neq i}d_jZ(z,\epsilon_j)\left(\Lambda(z)-\Lambda(\epsilon_j)\right) \nonumber\\
&+&\sum_{\alpha}Z(z,x_{\alpha})\left[Z(z,x_{\alpha})-2d_i Z(\epsilon_i,x_{\alpha})\right],
\end{eqnarray}
which holds if $z \neq \epsilon_j$ for all $j\neq i$. The evaluation of these equations at $z=\epsilon_i$ results in the known set of equations for $\Lambda_i$ if $d_i=1/2$, but for larger degeneracies the set of equations also depends on
\begin{equation}
\Lambda_{i}^{(2)}\equiv\sum_{\alpha=1}^N Z_{i\alpha}^2.
\end{equation}
By taking the derivative of Eq. (\ref{deg:conteq}), additional equations can be obtained linking these higher-order variables to the original variables. The derivation of this method is given in Appendix \ref{appendixB}. It is remarkable that the derivative of $\Lambda(z)$, a summation over $Z(z,x_{\alpha})$, can still be related to $\Lambda^{(2)}(z)$, a summation over $Z(z,x_{\alpha})^2$.

\begin{figure}[tb!]                      
 \begin{center}
 \includegraphics[width=\columnwidth]{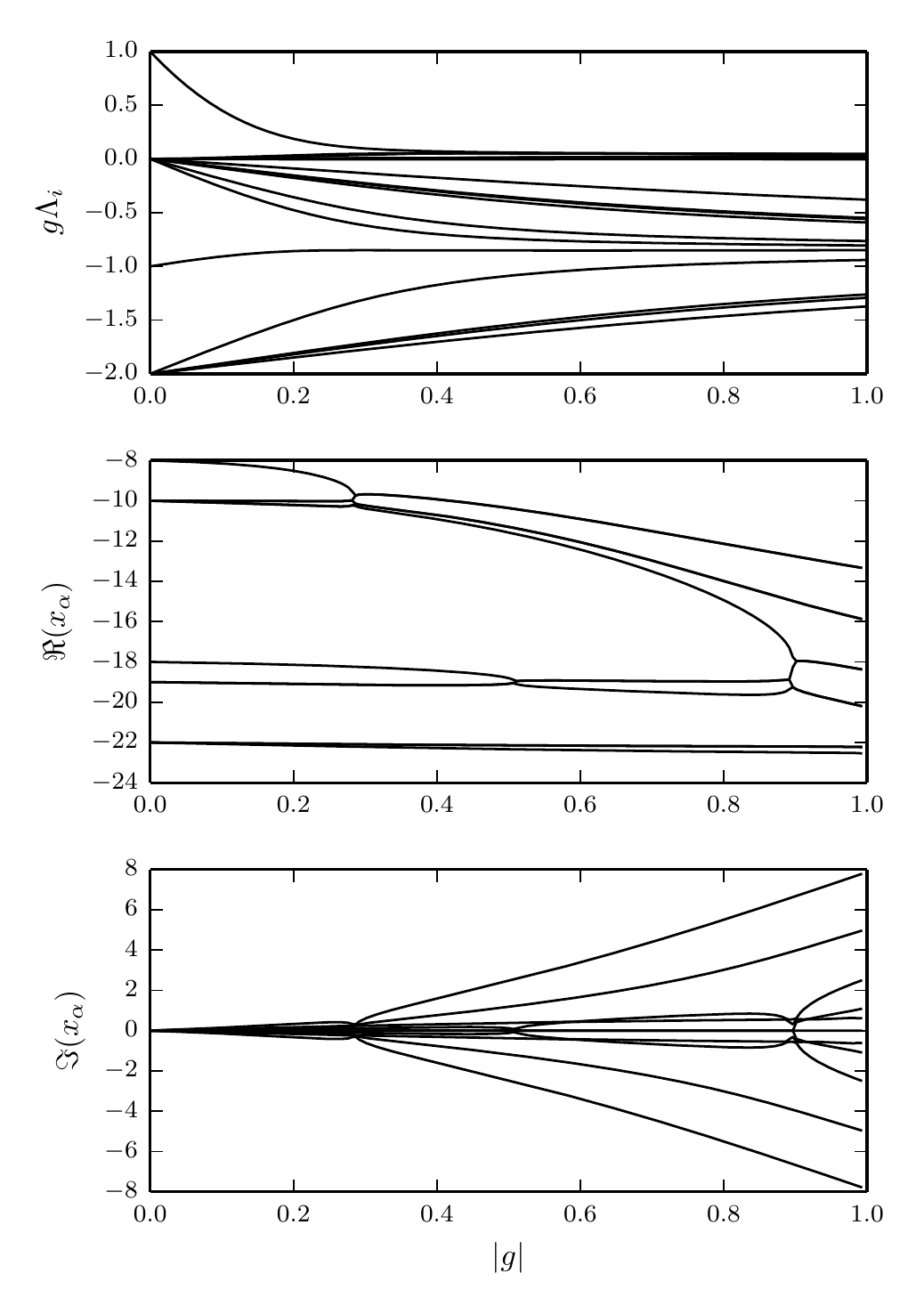}
 \caption{Evolution of the variables for the ground state containing 11 excitations in 10 levels for a varying (negative) coupling constant $g$, describing the pairing of neutrons in $^{56}$Fe by means of a level-independent reduced BCS model. The parameters of the system are taken from [\onlinecite{rombouts_solving_2004}], where the set of $d_i$ are given by $[3/2,1,1/2,2,1,3/2,1/2,1/2,1,5/2]$ with a maximal degeneracy of $2d_i+1=6$. The first figure shows the evolution of the eigenvalue-based variables, while the second and third figure correspond to the real respectively the imaginary part of the rapidities.  \label{fig:Stefan}}
 \end{center}
\end{figure}

As an example, the evaluation of the first derivative results in 
\begin{eqnarray}\label{deg:firstder}
\Lambda_i\left(N\Gamma+\Lambda_{i}^{(2)}\right)&=&-\frac{1}{g}\left(N\Gamma+\Lambda_{i}^{(2)}\right)\nonumber\\&&+\sum_{j \neq i}d_j(\Gamma+Z_{ij}^2)\left(\Lambda_i-\Lambda_j\right) \nonumber\\
&&+\sum_{j \neq i}d_j Z_{ij}\left(\Lambda_{i}^{(2)}+\Gamma\right)\nonumber\\
&& + \Gamma(1-d_i)\Lambda_i +(1-d_i)\Lambda_{i}^{(3)}.
\end{eqnarray}
For $d_i=1$, we have obtained a closed set of equations in $\{\Lambda_i,\Lambda_i^{(2)}\}$ for each level. For arbitrary $d_i$, the first $2d_i-1$ derivatives results in a closed set of equations. An additional equation for the total number of excitations can easily be determined as
\begin{equation}
N=-\sum_{i=1}^n g d_i \Lambda_i.
\end{equation}
As an illustration, the results for a level-independent reduced BCS model describing neutron superfluidity\cite{rombouts_solving_2004} in $^{56}$Fe  are given in Figure \ref{fig:Stefan}. The energy levels are determined as the levels of a spherical symmetrical Woods-Saxon potential and the $d_i$, dependent on the angular momentum quantum numbers, vary from $1/2$ to $5/2$.

\section{Form factors}

\subsection{Overlap with Slater determinants}
When performing calculations with Bethe Ansatz states, it is customary to expand them in a basis set of uncorrelated wavefunctions. These expansion coefficients are given by permanents of matrices in the general case, following from the theory of multivariate generating functions \cite{percus_combinatorial_1971}. However, these expressions are not practical for computational purposes since the evaluation of a permanent scales exponentially with the matrix size. Compared to the determinant, which can be evaluated in a polynomial time, permanents require a factorial scaling computational time and therefore severely limit the size of the systems that can be considered.

Luckily, multiple formulae exist for the rational model linking the permanents to determinants, allowing numerically efficient expansions of the Bethe Ansatz state. Here, we present several such results for $d_i=1/2$ XXZ systems as a generalization of the results previously obtained for the XXX model \cite{faribault_determinant_2012} and the inhomogeneous Dicke model \cite{tschirhart_algebraic_2014}. Starting from these expressions, determinant expressions for systems with arbitrary $d_i$ can also be obtained as a limiting case where several rapidities coincide, but this will not be discussed here.

Multiple determinant formulae exist for the overlap of a Bethe Ansatz state in the rational model with a Slater determinant 
\begin{equation}
\ket{\{i_a\}}=\ket{\{i_1\dots i_N\}}=\prod_{a=1}^N S^{\dagger}_{i_a}\ket{\downarrow \dots \downarrow},
\end{equation}
which is an uncorrelated wavefunction and an eigenstate in the uncoupled limit ($g=0$). The Bethe Ansatz state (\ref{rg:bas}) can be expanded in this basis as
\begin{eqnarray}
\ket{\psi_N}&=&\prod_{\alpha=1}^N\left(\sum_{i=1}^n X_{i\alpha} S^{\dagger}_i\right)\ket{\theta} \nonumber \\
&=&\sum_{\{i_a\}}\ket{\{i_a\}}\braket{\{i_a\}|\psi_N} \\
&=&\sum_{\{i_a\}}\ket{\{i_a\}} \perm \left(X_{i_a \alpha}\right),
\end{eqnarray}
where $\{i_a\}$ runs over the partitioning of $n$ levels over $N$ different excitations modes and
\begin{equation}
\braket{\{i_a\}|\psi_N}= \perm \left(\left[X_{i_a \alpha}\right]\right),
\end{equation}
which is the permanent of an $(N \times N)$ matrix with matrix elements given by $X_{i_a \alpha}$. This is a formula independent of the integrability or the explicit expression for $X$ and is simply a result of the structure of the wavefunction. However, by imposing the Gaudin algebra on the matrix elements, it is always possible to rewrite this permanent as a determinant. 
\begin{eqnarray}
&&\perm \left(\left[X_{i_a \alpha}\right]\right)\nonumber\\
&&\ \ =\frac{\left(\prod_{a>b}X_{i_ai_b}\right)\left(\prod_{\beta>\alpha}X_{\alpha\beta}\right)}{\prod_{a,\alpha}X_{i_a\alpha}}\det\left[X_{i_a\alpha}*X_{i_a\alpha}\right]
\end{eqnarray}
where the Hadamard product of two matrices is introduced, defined as $(A*B)_{ij}=A_{ij}B_{ij}$. This is a generalization of the Borchardt or Izergin determinant representation for the rational model \cite{borchardt_bestimmung_1857,izergin_determinant_1992,singer_bijective_2004} and has a similar structure to some results presented previously for the trigonometric model\cite{hao_determinant_2012}.

In ref. [\onlinecite{faribault_determinant_2012}], a case was made for a RG theory that would require only the eigenvalue-based variables instead of the rapidities. This would be desirable from a numerical point of view as the eigenvalue-based variables are free of singularities, opposed to the singularity-prone rapidities. In addition, this would enable us to skip the inversion step, which is the main bottleneck of this method. 

For the rational model, it was shown by Faribault and Schuricht \cite{faribault_determinant_2012} that 
\begin{equation}
\perm\left(\left[\frac{1}{\epsilon_{i_a}-x_{\alpha}}\right]\right)=\det J
\end{equation}
where the left-hand size is the permanent of an $N \times N$ Cauchy matrix and the right hand side is the determinant of a matrix $J$ defined as
\begin{equation}\label{rat:overlap}
J_{ab}=\begin{cases}\sum_{\alpha=1}^N\frac{1}{\epsilon_{i_a}-x_{\alpha}}-\sum_{c \neq a}^N\frac{1}{\epsilon_{i_a}-\epsilon_{i_c}}  &\text{if } a=b\\
\frac{1}{\epsilon_{i_a}-\epsilon_{i_b}}  &\text{if } a \neq b
\end{cases}
\end{equation}
where the only dependence on the rapidities is through the diagonal elements, which can be expressed in terms of the eigenvalue-based variables. The ingenious proof of this expression is based on a recursion relation which was found to hold for both sides of Eq. (\ref{rat:overlap}). We will show that these results can be easily generalized to XXZ models.

The Gaudin equations can be used to rewrite the permanent in the XXZ model in a structure similar to a Cauchy matrix. We once again introduce an auxiliary level $r$ and use the Gaudin equations to write
\begin{equation}
X_{i\alpha}=\frac{X_{ri}X_{r\alpha}}{Z_{ri}-Z_{r\alpha}}, \qquad Z_{i\alpha}=\frac{\Gamma+Z_{ri}Z_{r \alpha}}{Z_{ri}-Z_{r\alpha}}.
\end{equation}
Rewriting all matrix elements in the permanent $\phi$ allows us to take
\begin{eqnarray}
&&\perm\left(\left[X_{i_a \alpha}\right]\right)=\perm\left(\left[\frac{X_{ri_a}X_{r\alpha}}{Z_{ri_a}-Z_{r\alpha}}\right]\right)\nonumber\\
&&\ =\left(\prod_a X_{ri_a}\right)\left( \prod_{\alpha} X_{r\alpha}\right) \perm\left(\left[\frac{1}{Z_{ri_a}-Z_{r\alpha}}\right]\right),
\end{eqnarray}
resulting in a Cauchy matrix, which we can rewrite as a determinant
\begin{equation}
\perm\left(\left[X_{i_a \alpha}\right]\right)= \left(\prod_a X_{ri_a}\right)\left( \prod_{\alpha} X_{r\alpha}\right) \det J
\end{equation}
with $J$ redefined as
\begin{equation}
J_{ab}=
\begin{cases}
\sum_{\alpha=1}^N\frac{1}{Z_{ri_a}-Z_{r\alpha}}-\sum_{c \neq a}^N\frac{1}{Z_{ri_a}-Z_{ri_c}}  &\text{if } a=b\\
\frac{1}{Z_{ri_a}-Z_{ri_b}}  &\text{if } a \neq b\\
\end{cases}
\end{equation}
By multiplying each row and column $c$ with $X_{ri_c}$ and compensating for these factors in the prefactor, this can be written as
\begin{equation}
\perm\left(\left[X_{i_a \alpha}\right]\right)=\frac{\prod_{\alpha}X_{r\alpha}}{\prod_{a}X_{ri_a}} \det J
\end{equation}
with
\begin{equation}
J_{ab}=\begin{cases}
\sum_{\alpha=1}^N\frac{X_{ri_a}^2}{Z_{ri_a}-Z_{r\alpha}}-\sum_{c \neq a}^N\frac{X_{ri_a}^2}{Z_{ri_a}-Z_{ri_c}}  &\text{if } a=b\\
\frac{X_{ri_a}X_{ri_b}}{Z_{ri_a}-Z_{ri_b}}  &\text{if } a \neq b,
\end{cases}
\end{equation}
where we recognize $X_{i_ai_b}$ in the off-diagonal elements and the diagonal elements can be rewritten by using the Gaudin algebra until
\begin{equation}
J_{ab}=\begin{cases}\sum_{\alpha=1}^NZ_{i_a\alpha}-\sum_{c \neq a}^NZ_{i_ai_c}+Z_{ri_a}  &\text{if } a=b\\
X_{i_ai_b}  &\text{if } a \neq b
\end{cases},
\end{equation}
The only explicit dependency on the rapidities is now found in the prefactor and in the diagonal elements, which only depend on the eigenvalue-based variables.  Fortunately, the prefactor can be absorbed in the definition of the state without loss of generality, which will still allow the calculation of all necessary normalizations and prefactors. Note the appearance of $Z_{ri}$ in the diagonal elements, which can be easily evaluated and links the matrix to the prefactor for the Bethe Ansatz state through our choice of the additional level $r$. 

To recapitulate, the overlap of a state
\begin{equation}
\ket{\{x_{\alpha}\}}=\prod_{\alpha=1}^N\left(\sum_{i=1}^n \frac{X_{i\alpha}}{X_{r\alpha}} S^{\dagger}_i\right)\ket{\downarrow \dots \downarrow}
\end{equation}
with a Slater determinant
\begin{equation}
\ket{\{i_a\}}=\ket{\{i_1\dots i_N\}}=\prod_{j=1}^N S^{\dagger}_{i_j}\ket{\downarrow \dots \downarrow}
\end{equation}
is given by 
\begin{equation}\label{ov:XXZ}
\braket{\{i_1\dots i_N\}|\{x_{\alpha}\}}=\frac{\det J}{\prod_{a}X_{r i_a}} 
\end{equation}
with
\begin{equation}
J_{ab}=\begin{cases}
\Lambda_{i_a}-\sum_{c \neq a}^NZ_{i_ai_c}+Z_{ri_a}  &\text{if } a=b\\
X_{i_ai_b}  &\text{if } a \neq b
\end{cases}.
\end{equation}

\subsection{Dual representations}
So far, all Bethe Ansatz states have been created by acting with creation operators on the vacuum state, destroyed by all annihilation operators. Due to the particle-hole symmetry of the RG models, every Bethe Ansatz state can also be constructed by acting on the fully-filled state, annihilated by all creation operators, with a set of generalized annihilation operators, the so-called dual representation. 

A renormalized Bethe Ansatz state is given in its normal and dual representation as
\begin{eqnarray}
\ket{\{x_{\alpha}\}}&=&\prod_{\alpha=1}^N\left(\sum_{i=1}^n \frac{X_{i\alpha}}{X_{r\alpha}} S^{\dagger}_i\right)\ket{\downarrow \dots \downarrow}, \\
\ket{\{x_{\alpha'}\}}&=&\prod_{\alpha'=1}^{n-N}\left(\sum_{i=1}^n \frac{X_{i\alpha'}}{X_{r\alpha'}} S_i\right)\ket{\uparrow \dots \uparrow}.
\end{eqnarray}
The RG equations for the dual state are given by
\begin{equation}
-1+\frac{g}{2}\sum_{i}Z_{j\alpha'} = g \sum_{\beta' \neq \alpha'}^{n-N}Z_{\beta'\alpha'}, \qquad \forall \alpha'=1 \dots n-N
\end{equation}
or, written in the eigenvalue-based variables,
\begin{equation}
\left[\Lambda_i'\right]^2=N(n-N)\Gamma+\frac{2}{g}\Lambda_i'+\sum_{j \neq i} Z_{ji}(\Lambda_j'-\Lambda_i'), \ \ \forall i=1\dots n.
\end{equation}
Both representations describe, up to the normalization, the same state. So by comparing the eigenvalues of the constants of motion
\begin{eqnarray}
r_i&=&\frac{1}{2}\left(-1-g\Lambda_i +\frac{g}{2}\sum_{k\neq i}^nZ_{ik}\right) \nonumber\\
&=&\frac{1}{2}\left(1-g\Lambda_i'+\frac{g}{2}\sum_{k\neq i}^n Z_{ik}\right),
\end{eqnarray}
it can be seen that the dual representation of an eigenstate is determined by a set of dual eigenvalue-based variables given by
\begin{equation}
\Lambda_i'=\Lambda_i+\frac{2}{g}, \qquad \forall i=1 \dots n,
\end{equation}
which can be verified by simply substituting these variables in the dual eigenvalue-based equations.
The existence of the dual state allows one to write the overlap of a normal state and a dual state as the overlap of a Bethe Ansatz state with the fully-filled state. This will be exploited in the following section to obtain the normalization of the Bethe Ansatz states and several form factors.

\subsection{Normalization and form factors}
Due to the dual representation, we have two different representations of each Bethe Ansatz state. These two states have different normalizations and can be written as 
\begin{eqnarray}
\ket{\{x_\alpha\}}&=&\prod_{\alpha=1}^N \left(\sum_{i=1}^n \frac{X_{i\alpha}}{X_{r\alpha}}S^{\dagger}_i\right)\ket{\downarrow \dots \downarrow} = N_{\alpha} \ket{\{x_\alpha\}}_n \nonumber \\
\ket{\{x_\alpha'\}}&=&\prod_{\alpha'=1}^{n-N} \left(\sum_{i=1}^n \frac{X_{i\alpha'}}{X_{r\alpha'}}S_i\right)\ket{\uparrow \dots \uparrow} = N_{\alpha'} \ket{\{x_\alpha\}}_n \nonumber
\end{eqnarray}
where $\ket{\{x_{\alpha}\}}_n$ is the normalized eigenstate. This can be used to calculate
\begin{eqnarray}
&&\braket{\{x_{\alpha'}\}|\{x_{\alpha}\}}=N_{\alpha}N_{\alpha'}\nonumber\\
&&\ \ =\braket{\uparrow \dots \uparrow |\prod_{\alpha'=1}^{n-N} \left(\sum_{i=1}^n \frac{X_{i\alpha'}}{X_{r\alpha'}}S^{\dagger}_i\right)\prod_{\alpha=1}^N \left(\sum_{i=1}^n \frac{X_{i\alpha}}{X_{r\alpha}}S^{\dagger}_i\right) |\downarrow \dots \downarrow} \nonumber\\
&&\ \ =\frac{\det J}{\prod_{i=1}^n X_{ri}} 
\end{eqnarray}
with
\begin{equation}
J_{ij}=\begin{cases}
2\Lambda_{i}+\frac{2}{g}-\sum_{k \neq i}^nZ_{ik}+Z_{ri}  &\text{if } i=j\\
X_{ij}  &\text{if } i \neq j
\end{cases},
\end{equation}
since the overlap between the two states is just the overlap of a Bethe Ansatz state determined by $\{x\}=\{x_{\alpha}\}\cup\{x_{\alpha'}\}$ or by $\Lambda_i=\Lambda_i+\Lambda_i'=2\Lambda_i+2/g$ with $\ket{\uparrow \dots \uparrow}$.
The ratio $N_{\alpha}/N_{\alpha'}$ can easily be found by taking the overlap of both representations with a single reference state and taking this ratio. Knowing both the ratio and the product of the normalizations, the normalization of both representations is uniquely defined

In a similar manner as for the XXX model \cite{faribault_determinant_2012,tschirhart_algebraic_2014}, these results can be used to calculate form factors. Since this is a generalization of the results presented in these papers, we will only summarize the results for the XXZ model here. 

The form factor for the raising operator between a state $\{x_\alpha\}$ with $N$ excitations and a dual state  $\{x_\mu'\}$ with $N+1$ excitations is given by
\begin{equation}
\braket{\{x_\mu'\}|S^{\dagger}_k|\{x_\alpha\}}=\frac{\det J^k}{\prod_{i \neq k} X_{ri}}
\end{equation}
with
\begin{equation}
J^k_{ab}=\begin{cases}
\Lambda_{i}^{\alpha}+\Lambda_{i}^{\mu}+\frac{2}{g}-\sum_{l \neq i,l \neq k}^nZ_{il}+Z_{ri}  &\text{if } a=b\\
X_{ab}  &\text{if } a \neq b
\end{cases}
\end{equation}
and $a,b \neq k$. The form factor for a lowering operator is then given by the Hermitian conjugate of this expression.

The form factor for local operators $S_k^0$ can similarly be found, both for diagonal and off-diagonal expectation values. Here we will make the dependency on the coupling constant explicit and define $\{x_\alpha(g)\}$ as an eigenstate at coupling constant $g$.

From the Hellmann-Feynman theorem we obtain for the diagonal expectation values
\begin{eqnarray}
&&\braket{\{x_\alpha'(g)\}|S^0_k|\{x_\alpha(g)\}}=\nonumber\\
&&\qquad\frac{1}{2}\left(-1+g^2\frac{\partial \Lambda^{\alpha}_k}{\partial g}\right)\braket{\{x_\alpha'(g)\}|\{x_\alpha(g)\}}
\end{eqnarray}
while the off-diagonal expectation values can be found as
\begin{eqnarray}
&&\braket{\{x_\mu'(g)\}|S^0_k|\{x_\alpha(g)\}} \nonumber\\
&&\qquad=\frac{1}{2}\left(g\Lambda^{\mu}_k-g\Lambda^{\alpha}_k+2\right)\frac{\braket{\{x_\mu'(g)\}|\{x_\alpha(g+dg)\}}}{dg} \nonumber\\
&&\qquad=\frac{1}{2}\left(g\Lambda^{\mu}_k-g\Lambda^{\alpha}_k+2\right)\sum_{k=1}^n\frac{\partial \Lambda^{\alpha}_k}{\partial g} \det \tilde{J}^k
\end{eqnarray}
with
\begin{equation}
\tilde{J}^k_{ab}=
\begin{cases}
\Lambda_{i}^{\alpha}+\Lambda_{i}^{\mu}+\frac{2}{g}-\sum_{l \neq i}^nZ_{il}+Z_{ri}  &\text{if } a=b\\
X_{ab}  &\text{if } a \neq b
\end{cases}
\end{equation}
and $a,b \neq k$. These form factors can be written in an expression only dependent on the eigenvalue-based variables, so the rapidities do not need to be determined explicitly. However, if these have been determined, other form factors can also be calculated by commuting the operator through the creation operators in the Bethe Ansatz state, as has been shown in [\onlinecite{zhou_superconducting_2002}]. This allows the computation of form factors such as e.g. $S^{\dagger}_k S_l$ as a sum over determinants. 

It is interesting to note that the expectation value for $S_i^0$ offers a physical interpretation for $\Lambda_i$, since the derivative of $\Lambda_i$ to $g$ fully determines the filling of the level $i$ as described by $\braket{S_i^0}$. Knowledge of the evolution of $\Lambda_i$ with a changing coupling constant is then equivalent to knowing how the $N$ excitations are distributed over the $n$ levels.

\section{Conclusions and outlook}
In the present paper, we presented a numerical solution method for the RG equations for XXZ integrable models, while also proposing determinant expressions for the normalization and form factors of the Bethe Ansatz states that are only dependent on the new set of eigenvalue-based variables. The availability of an efficient solution method and explicit expressions for the overlaps opens up possibilities for the investigation of integrable and non-integrable quantum systems.

An efficient numerical expression for overlaps between two Bethe Ansatz states allows for a numerically nearly exact investigation of the dynamics resulting from a quantum quench in the reduced BCS pairing model \cite{faribault_quantum_2009}, while for models close to integrability the eigenstates of the integrable model are being used to study decoherence \cite{van_den_berg_competing_2014}. These results were all obtained for the rational XXX model, and it would be interesting to investigate similar dynamics for XXZ models, where a rich phase diagram is known \cite{rombouts_quantum_2010}. 

Due to the favourable computational scaling of this method it is also possible to numerically approach the thermodynamic limit and investigate the limiting behaviour of the system. Such an investigation has been carried out for the reduced BCS Hamiltonian by El Araby and Baeriswyl \cite{el_araby_order_2014}, where exact results for large system sizes were compared with the BCS Ansatz. Again, it would be worthwhile to compare the results rigourously with the results from BCS mean-field theory \cite{ibanez_exactly_2009}.

The overlaps between Bethe Ansatz states and Slater determinants have also been used in quantum chemistry for strongly correlated systems. These can not be accurately described by means of a single Slater determinant, and several wavefunctions have been proposed to capture the correlation present in the system. One example is the class of geminal-based wavefunctions, which have a structure highly reminiscent of the Bethe Ansatz states found in the Gaudin models. Unfortunately, these wavefunctions are only computationally feasible if a numerical efficient expression is known for the overlaps with Slater determinants. One such class (APr2G) is based on the rational XXX model \cite{johnson_size-consistent_2013,limacher_new_2013,tecmer_assessing_2014}, and the presented results should allow for an extension of these geminal-based wavefunctions with a class based on the XXZ models. The results presented here should also allow for a variational approach based on the Bethe Ansatz states in the XXZ model. By starting from Richardson's general solution for the Gaudin algebra (Eq. \ref{ga:richalg}) and varying over the free parameters in the wavefunction, it is possible to scan over all possible Gaudin models (XXX and XXZ), allowing more freedom and as such a better approximation to the energy compared to a variational approach based solely on the XXX model.

It was already noted by Babelon and Talalaev that similar equations could be found for the XXX Heisenberg spin chain \cite{babelon_bethe_2007}, so the question naturally arises if this method can be generalized to other Bethe Ansatz equations obtained from the quantum inverse scattering method \cite{korepin_quantum_1993}. 

\section*{Acknowledgements}
PWC and SDB acknowledge financial support from FWO-Vlaanderen as pre-doctoral and post-doctoral fellows respectively.
\appendix
\section{Deriving the equations}
\label{appendixA}
In this appendix we derive the equations for the eigenvalue-based variables by making use of the Richardson-Gaudin equations and the Gaudin algebra. All summations with  indices labelled by Greek characters run over the $N$ rapidities, while for indices labeled by Latin characters the summations run over the $n$ energy levels. Starting from the definition of the variable $\Lambda_i$, we can write
\begin{eqnarray}
\Lambda_i^2&=&\sum_{\alpha,\beta}Z_{i\alpha}Z_{i\beta}=\sum_{\alpha,\beta \neq \alpha}Z_{i\alpha}Z_{i\beta}+\sum_{\alpha}Z_{i\alpha}^2\nonumber\\
&=&-\sum_{\alpha,\beta \neq \alpha}\left[\Gamma+Z_{\alpha\beta}(Z_{\alpha i}+Z_{i\beta})\right]+\sum_{\alpha}Z_{i\alpha}^2,
\end{eqnarray}
where we have used Eq. \ref{ga:Zeq} to rewrite $Z_{i\alpha}Z_{i\beta}$. The antisymmetry of $Z_{i\beta}$ and $Z_{\alpha \beta}$ can be used to obtain
\begin{eqnarray}
\Lambda_i^2&=&-\sum_{\alpha,\beta \neq \alpha}\left(\Gamma+2Z_{\alpha\beta}Z_{\alpha i}\right)+\sum_{\alpha}Z_{i\alpha}^2 \nonumber\\
&=&-N(N-1)\Gamma+2\sum_{\alpha }Z_{\alpha i}\left(\sum_{\beta \neq \alpha}Z_{\beta \alpha}\right)+\sum_{\alpha}Z_{i\alpha}^2 \nonumber
\end{eqnarray}
Making use of the RG equations (Eq. \ref{ga:RGeq}), we obtain
\begin{eqnarray}
\Lambda_i^2&=&-N(N-1)\Gamma+2\sum_{\alpha }Z_{\alpha i}\left(\frac{1}{g}+\frac{1}{2}\sum_{j}Z_{j\alpha}\right)+\sum_{\alpha}Z_{i\alpha}^2 \nonumber\\
&=&-N(N-1)\Gamma+\frac{2}{g}\sum_{\alpha }Z_{\alpha i}+\sum_{\alpha}\sum_{j \neq i} Z_{\alpha i} Z_{j \alpha},
\end{eqnarray}
which can again be rewritten by making use of Eq. \ref{ga:Zeq}.
\begin{eqnarray}
\Lambda_i^2 &=&-N(N-1)\Gamma+\frac{2}{g}\sum_{\alpha }Z_{\alpha i} \nonumber\\
&&\qquad+\sum_{\alpha}\sum_{j \neq i} \left(\Gamma+Z_{ji}(Z_{j\alpha}+Z_{\alpha_i})\right) \nonumber\\
&=&N(n-N)\Gamma-\frac{2}{g}\Lambda_i+\sum_{j \neq i} Z_{ji}(\Lambda_j-\Lambda_i),
\end{eqnarray}
resulting in a set of equations closed in the variables $\Lambda_i, i=1 \dots n$.

\section{Obtaining equations for degenerate models}
\label{appendixB}
Here, we show how to obtain additional equations for degenerate models starting from a continuous equation for the eigenvalue-based variables. We define
\begin{equation}
\Lambda_{i}^{(p)}\equiv\sum_{\alpha=1}^N Z_{i\alpha}^p, \qquad p \in \mathbb{N}
\end{equation}
and 
\begin{equation}
\Lambda^{(p)}(z)\equiv\sum_{\alpha=1}^N Z(z,x_{\alpha})^p, \qquad p \in \mathbb{N}.
\end{equation}
By taking the derivative with respect to $z$ of Eq. (\ref{deg:conteq}), additional equations can be found. Unfortunately, an explicit expression for $Z(z,\epsilon_i)$ is needed if we want to relate the derivative of $\Lambda(z)$ to $\Lambda^{(2)}(z)$ and obtain a closed set of equations. In order to circumvent this problem, we introduce a fixed auxiliary level $\epsilon_r$, so that every $Z$ can now again be parametrized as in Eq. (\ref{ga:Zeq}),
\begin{equation}
Z(z,x_{\alpha})=\frac{Z(\epsilon_r,z)Z(\epsilon_r,x_{\alpha})+\Gamma}{Z(\epsilon_r,z)-Z(\epsilon_r,x_{\alpha})},
\end{equation}
so instead of deriving this equation with respect to the continuous variable $z$, we can derive with respect to $Z(\epsilon_r,z)$ and multiply with $-X_{rz}^2$, resulting in an equation independent of the explicit expression for $Z$. For notational ease, we take $Z_{rz}=Z(\epsilon_r,z)$. Then
\begin{equation}
-X_{rz}^2 \frac{d}{d Z_{rz}} Z(z,x_{\alpha})=X(z,x_{\alpha})^2=Z(z,x_{\alpha})^2+\Gamma,
\end{equation}
which is independent of the auxiliary level. The action of this operator (derivation and multiplication) is given by
\begin{equation}
\Lambda^{(p)}(z) \to p\Lambda^{(p+1)}(z)+p\Gamma\Lambda^{(p-1)}(z),
\end{equation}
and
\begin{equation}
Z(z,\epsilon_j)^p \to p Z(z,\epsilon_j)^{(p+1)}+p\Gamma Z(z,\epsilon_j)^{(p-1)}.
\end{equation}
By acting with this operator on the equations for $\Lambda(z)$, we obtain
\begin{eqnarray}
\Lambda(z)(N\Gamma&&+\Lambda^{(2)}(z))=-\frac{1}{g}\left(N\Gamma+\Lambda^{(2)}(z)\right)\nonumber\\
&&+\sum_{j \neq i}d_j(\Gamma+Z(z,\epsilon_j)^2)\left(\Lambda(z)-\Lambda(\epsilon_j)\right) \nonumber\\
&&+\sum_{j \neq i}d_jZ(z,\epsilon_j)\left(\Lambda^{(2)}(z)+\Gamma\right)\nonumber\\ 
&&+\sum_{\alpha}Z(z,x_{\alpha})^2\left(Z(z,x_{\alpha})-d_i Z(\epsilon_i,x_{\alpha})\right)\nonumber\\
&&+\Gamma\left(\Lambda(z)-d_i \Lambda(\epsilon_i)\right) 
\end{eqnarray}
resulting in Eq. (\ref{deg:firstder}) when evaluated in $z=\epsilon_i$. Successive derivatives result in as many additional equations as there are variables, allowing a closed set of equations to be obtained.

\bibliography{MyLibrary}

\end{document}